\documentclass[
aip,longbibliography,
 amsmath,amssymb,
 preprint,
]{revtex4-1}
\usepackage{graphicx}
\usepackage{dcolumn}
\usepackage{bm}
\usepackage{lipsum}
\usepackage[mathlines]{lineno}
\usepackage[utf8]{inputenc}
\usepackage{hyperref} 
\usepackage[T1]{fontenc}
\usepackage{mathptmx}
\usepackage{etoolbox}

\makeatletter
\def\@email#1#2{%
 \endgroup
 \patchcmd{\titleblock@produce}
  {\frontmatter@RRAPformat}
  {\frontmatter@RRAPformat{\produce@RRAP{*#1\href{mailto:#2}{#2}}}\frontmatter@RRAPformat}
  {}{}
}%
\makeatother

\begin{document}

\title[Free energy and rates of transmembrane dimerization in lipid bilayers]{Free energy, rates, and mechanism of transmembrane dimerization in lipid bilayers from dynamically unbiased molecular dynamics simulations}
\author{Emil Jackel}
\thanks{Equal contributions}
\affiliation{Institute of Biophysics, Goethe University Frankfurt, Frankfurt am Main, Germany.}
\affiliation{Frankfurt Institute for Advanced Studies, Frankfurt am Main, Germany.}
\author{Gianmarco Lazzeri}
\thanks{Equal contributions}
\affiliation{Institute of Biochemistry, Goethe University Frankfurt, Frankfurt am Main, Germany.}
\affiliation{Frankfurt Institute for Advanced Studies, Frankfurt am Main, Germany.}
\author{Roberto Covino*}
\affiliation{Institute of Computer Science, Goethe University Frankfurt, Frankfurt am Main, Germany.}
\affiliation{Frankfurt Institute for Advanced Studies, Frankfurt am Main, Germany.}
\email[Author to whom any correspondence should be addressed: ]{covino@fias.uni-frankfurt.de}
\date{\today}

\begin{abstract}
The assembly of proteins in membranes plays a key role in many crucial cellular pathways. Despite their importance, characterizing transmembrane assembly remains challenging for experiments and simulations. Equilibrium molecular dynamics simulations do not cover the time scales required to sample the typical transmembrane assembly. Hence, most studies rely on enhanced sampling schemes that steer the dynamics of transmembrane proteins along a collective variable that should encode all slow degrees of freedom. However, given the complexity of the condensed-phase lipid environment, this is far from trivial, with the consequence that free energy profiles of dimerization can be poorly converged. Here, we introduce an alternative approach, which relies only on simulating short, dynamically unbiased trajectory segments, avoiding using collective variables or biasing forces. By merging all trajectories, we obtain free energy profiles, rates, and mechanisms of transmembrane dimerization with the same set of simulations. We showcase our algorithm by sampling the spontaneous association and dissociation of a transmembrane protein in a lipid bilayer, the popular coarse-grained Martini force field. Our algorithm represents a promising way to investigate assembly processes in biologically relevant membranes, overcoming some of the challenges of conventional methods. 
\end{abstract}

\maketitle

\section{\label{sec:introduction}Introduction}

In cellular membranes, thousands of different lipids and proteins jointly self-organize in spatially heterogeneous and temporally dynamic structures, which carry out key regulatory functions in the cell. The basic structural elements in a large class of membrane proteins are transmembrane helices (TMHs), alpha helices that span the width of the lipid bilayers in which they are inserted~\cite{cournia_membrane_2015}. Membrane proteins can be tethered to the bilayer via a single TMH, like in the case of receptor tyrosine kinases on cellular surfaces~\cite{lemmon2010cell}, or have a fold composed of multiple TMHs, like, for example, g-coupled protein receptors~\cite{zhang_g_2024}. 

Understanding how TMHs assemble into dimers is key to gaining a mechanistic understanding of many fundamental cellular processes. For instance, receptor tyrosine kinases form dimers in the plasma membrane in response to external triggers as ligand binding to activate downstream pathways~\cite{lemmon2010cell}. In the Endoplasmic Reticulum, TMHs of Ire1 assemble into oligomers and large clusters in response to specific conditions that activate the unfolded protein response, a fundamental stress-response pathway in all eukaryotic cells~\cite{Walter2012,Covino2018,vath_cysteine_2021}. Additionally, folding multi-helical membrane proteins requires the sequential association of individual TMHs~\cite{hong_untangling_2022}. We still lack a quantitative understanding of transmembrane dimerization, particularly how the interplay between lipids and proteins controls it~\cite{corradi_emerging_2019,muller_characterization_2019}.

Outstanding experimental and simulation challenges limit the study of transmembrane assembly. Given their small size and the complexity of the membrane environment, very few structures of transmembrane dimers are known. Although undirect approaches are possible~\cite{Halbleib2017}, determining the thermodynamics and kinetics of membrane protein complexes in experiments is still extremely difficult~\cite{chadda_model-free_2018}. Molecular dynamics (MD) simulations became essential to investigate cellular membranes~\cite{muller_characterization_2019,corradi_emerging_2019,manna_understanding_2019,marrink_computational_2019}. 
Despite tremendous progress, equilibrium simulations showing the spontaneous and reversible assembly of membrane protein dimers and larger complexes are still mostly impossible~\cite{enkavi_multiscale_2019}. The time scales necessary to sample these events at atomistic resolution are inaccessible even to specialized supercomputers~\cite{pan_atomic-level_2019}. In the popular coarse-grained Martini force field, transmembrane dimers rapidly form, but sampling the spontaneous dissociation (a ``rare event'') is nearly impossible. This limitation, however, is not necessarily a problem of force field accuracy~\cite{domanski_balancing_2018, alessandri_pitfalls_2019,majumder_addressing_2021,claveras_cabezudo_scaling_2023}. The typical lifetime of a transmembrane dimer could be in the order of seconds or minutes, if not longer~\cite{chadda_model-free_2018}. 

Enhanced methods like umbrella sampling or metadynamics overcome this limitation by biasing the system's dynamics and steering it along the dimerization transition~\cite{henin2022enhanced}. The combination of umbrella sampling and statistical reweighing has provided fundamental insight into the interactions between membranes and proteins and became the de facto standard of free energy calculations for membrane proteins association~\cite{corey_insights_2019,souza2021martini}. However, umbrella sampling and related techniques do not directly provide kinetic information. Additionally, they require a collective variable (CV) to steer the system's dynamics. Crucially, the accuracy of the reconstructed free energy profile depends on how well the CV captures the mechanism of the dimerization process~\cite{peters2016reaction}. A CV that does not capture all relevant slow degrees of freedom will produce a non-converged and misleading free energy profile~\cite{domanski_convergence_2017}. Guessing good CVs for membrane protein assembly is particularly challenging because this process occurs in a condensed phase and requires quantifying collective many-body lipid reorganizations. This challenge is becoming more and more evident~\cite{aho_all_2024} and calling for sophisticated approaches that rely on increasing the dimensionality of the CVs to capture additional slow degrees of freedom~\cite{blazhynska_rigorous_2023,majumder_computing_2022}, extracting them from the data with machine-learning approaches~\cite{Chiavazzo2017,majumder_machine_2024}, or using a mixture of different biasing forces to facilitate convergence~\cite{ito_free-energy_2024}. 

Here, we present a method to obtain free energy, rates, and mechanisms of transmembrane dimerization from simulations without using biasing forces or needing good pre-defined CVs. We build on our recent machine learning-guided path sampling framework AIMMD~\cite{jung2023machine,lazzeri2023molecular}, which allows us to focus on simulating unbiased and reversible dimerization trajectories. Dynamically unbiased trajectories give direct access to the dimerization mechanism and facilitate the extraction of kinetic quantities. We show how AIMMD requires only the definition of bound and unbound states and provides accurate estimates of free energy, rates, and dimerization mechanisms at a moderate computational cost. 

\begin{figure*}
\centering
\makebox[\textwidth][c]{\includegraphics[width=.95\textwidth]{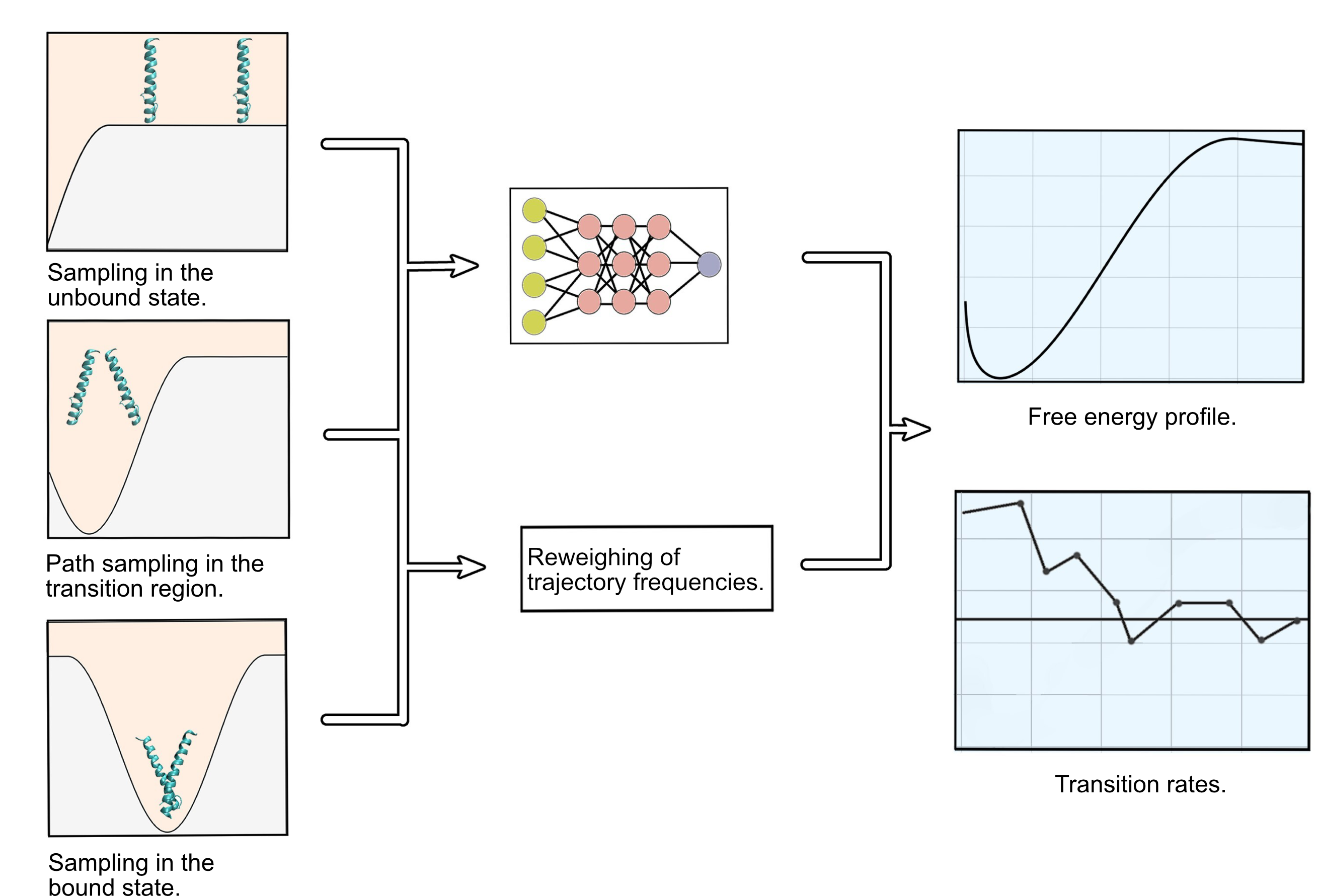}}
\caption{Schematic strategy of AIMMD for characterizing transmembrane protein dimerization processes. Equilibrium simulations are performed around the unbound and bound states, along with path-sampling simulations that enable the sampling in the transition region. Multiple independent workers are assigned to each task to speed up the process, making it embarrassingly parallel. A neural network  orchestrates the sampling of many short trajectories in an adaptive way. A reweighting algorithm is used to restore the frequencies of appearances of the individually simulated trajectory segments as if they came from  a long, unbiased equilibrium simulation. The algorithms provides estimates of the free energy profiles and transition rates.}
\label{fig:1}
\end{figure*}

\section{\label{sec:theory}Theory}

Our goal is to use AIMMD to sample and characterize efficiently the spontaneous and reversible $\mathrm{A}\rightleftharpoons\mathrm{B}$ transmembrane dimerization process. We summarize here the main elements of the AIMMD algorithm, while more details can be found in the original papers that introduced the method~\cite{jung2023machine,lazzeri2023molecular}. We will also highlight the differences in the implementation used to produce the results of this paper.

The goal is to sample the Path Ensemble (PE):
\begin{equation}
    \mathcal P[\textbf{x}] = \{\textbf{x}_1(t),~\dots,~\textbf{x}_n(t)\}~,
\end{equation}
a set of trajectory segments in configuration space $\textbf{x}_i(t)=\{\textbf{x}_i(0),~\dots,~\textbf{x}_i(L_i~\Delta t)\}$, saved at regular time intervals $\Delta t$ and of variable length $L_i=L[\textbf{x}_i]$. The PE can ideally be obtained by partitioning a long equilibrium trajectory describing a system that transitions between two metastable states $\mathrm{A}$ and $\mathrm{B}$. We will consistently use $\mathrm{A}$ for the bound state and $\mathrm{B}$ for the unbound state, and $\mathrm{R}$ for the reactive region in between where the dimer is assembling or disassembling. The long trajectory is divided into the following trajectory segments (Fig.~\ref{fig:1}):
\begin{itemize}
    \item[-] segments internal to $\mathrm{A}$---dynamics of the bound dimer;
    \item[-] segments internal to $\mathrm{B}$---dynamics of the unbound dimer;
    \item[-] excursions to $\mathrm{R}$ that start upon exiting $\mathrm{A}$ and end upon entering $\mathrm{A}$;
    \item[-] excursions to $\mathrm{R}$ that start upon exiting $\mathrm{B}$ and end upon entering $\mathrm{B}$;
    \item[-] transitions from $\mathrm{A}$ to $\mathrm{B}$---the dimer disassociates;
    \item[-] transitions from $\mathrm{B}$ to $\mathrm{A}$---the dimer associates.
\end{itemize}

Estimating free energy and rates requires sampling large excursions from the metastable states in $\mathrm{R}$ and entire transitions connecting the metastable states. 
In a long equilibrium simulation, these trajectory segments are very rare, which makes them usually impossible to sample by a straightforward brute-force approach. AIMMD is designed to enhance the sampling of significant excursions and transitions by initializing path sampling simulations in an informed manner.

We simulate trajectory segments in two ways:
\begin{itemize}
    \item[-] equilibrium simulations initialized at the boundary of states $\mathrm{A}$ and $\mathrm{B}$ mostly provide internal segments and small excursions that do not step that far from their origin states;
    \item[-] AIMMD path sampling simulations in the transition region provide large excursions and transitions.
\end{itemize}
Another advantage of sampling short trajectory segments is that we can dedicate many computing nodes to each of these tasks separately and simulate them in parallel. Parallel simulations are a significant improvement on the original algorithm, as they can leverage the power of high-performance computing clusters.

In order to improve the efficiency of AIMMD path sampling simulations and guarantee that the sampling converges to the PE, we control the selection of configurations from which we initialize simulations through $\lambda(x)$, a machine-learning (ML) approximation of the committor~\cite{jung2023machine}:
\begin{equation}
    p_{\mathrm{B}}(x) \approx \sigma(\lambda(x)),
\end{equation}
where $\sigma(\cdot)$ is the sigmoid function. $p_{\mathrm{B}}(x)$ is considered an optimal RC to describe thermally activated transitions between two states $\mathrm{A}$ and $\mathrm{B}$~\cite{peters2016reaction}. As such, it encodes the transition mechanism and defines the (equilibrium) transition state ensemble (TSE), where $p_{\mathrm{B}}=0.5$ and $\lambda(x)=0$. 

In AIMMD, the algorithm learns the committor on the fly by comparing the expectation of repeated path sampling simulations with their outcome~\cite{jung2023machine}. Each attempt to sample a transition relies on a ``two-way shooting'' move, i.e., the algorithm selects a configuration on a previous path (a shooting point) and initializes two simulations by using random velocities $\mathbf{v}$ and $\mathbf{-v}$, respectively. While repeatedly sampling, the algorithm trains the committor models by using a neural network (NN) that maximizes the likelihood of the two-way shooting results. By selecting starting configurations among the already simulated configurations and enforcing a uniform distribution along the committor, we quickly obtain a comprehensive sample of the PE, faster than uninformed path sampling algorithms. Compared to the original AIMMD algorithm, we directly select a new shooting point from the latest PE estimate; thus, at convergence, the shooting points follow the Boltzmann distribution restricted to their isocommittor surfaces. By allowing shooting points to come from all kinds of excursions and not just the latest accepted transition path sampling path, we foster exploration of the transition region.

By enhancing the sampling of large excursions in $\mathrm{R}$ and transitions, we increase their statistical weight. Thus, we must restore the equilibrium frequency through an importance-sampling reweighting procedure~\cite{lazzeri2023molecular,rogal2010reweighted}. The committor model is also crucial for an optimal reweighting of the PE trajectory segments. In substance, we associate to any segment $\textbf{x}_i$ a weight $w_i$ that recovers the frequency with which it would appear in a long equilibrium trajectory. Each excursion $\textbf{x}_i$ is characterized by its extreme RC value $\lambda_i^\star$, which quantifies how far it reaches from its origin state. Based on the equilibrium and path sampling simulations, we estimate the crossing probability $P_{\mathrm{A}}(\lambda)$---the probability that an equilibrium excursion from $\mathrm{A}$ ever crosses the RC value $\lambda$---and the  equivalent $P_{\mathrm{B}}(\lambda)$ from $\mathrm{B}$~\cite{rogal2010reweighted}. The algorithm  matches $P_{\mathrm{A}}(\lambda)$ and $P_{\mathrm{B}}(\lambda)$ with the weighted trajectories while also factoring in the RC values of the shooting points:
\begin{equation}
    w_i = \begin{cases}
    f_i~\dfrac{P_{\mathrm{A}}(\lambda_i^\star)}{m_{\mathrm{A}}(\lambda_i)}\quad&\text{if }\textbf{x}_i\text{ is an excursion from }\mathrm{A},\\
    f_i~\dfrac{P_{\mathrm{B}}(\lambda_i^\star)}{m_{\mathrm{B}}(\lambda_i)}~C\quad&\text{if }\textbf{x}_i\text{ is an excursion from }\mathrm{B},\\
    1/{n_{\text{in-}\mathrm{A}}}\quad&\text{if }\textbf{x}_i\text{ is internal to }\mathrm{A},\\
    C/{n_{\text{in-}\mathrm{B}}}\quad&\text{if }\textbf{x}_i\text{ is internal to }\mathrm{B},\\
    \end{cases}
\end{equation}
where $m_{\mathrm{A}}(\lambda_i^\star)$ and $m_{\mathrm{B}}(\lambda_i^\star)$ represent the unweighted statistics of the excursions at their extreme values, $f_i=1/n_{\mathrm{neigh},\mathrm{sp}}[\textbf{x}_i]$ is a prefactor that restores the equilibrium distribution of the two-way shooting paths around their shooting interface~\cite{falkner2023conditioning}, $n_{\text{in-}\mathrm{A}}$, $n_{\text{in-}\mathrm{B}}$ are the total number of internal to $\mathrm{A}$ and to $\mathrm{B}$ segments, and $C=P_{\mathrm{A}}(\mathrm{B}\equiv+\infty)/P_{\mathrm{B}}(\mathrm{A}\equiv-\infty)$ enforces the $\mathrm{A}\rightleftharpoons\mathrm{B}$ detailed balance in the reweighted segments. 

After reweighting, we obtain an estimate of the PE, which means that we can treat all configurations contained in all trajectory segments as coming from a long equilibrium simulation. The free energy profile along an arbitrary collective variable $q$ comes by projecting the PE~\cite{bolhuis2011relation}:
\begin{equation}
    F(q)~dq = -k_{\mathrm{B}}T\log\left[\sum_i w_i \sum_{t=1}^{L[\textbf{x}]-1}\delta\left(\textbf{x}_i(t\cdot\Delta t)-q(\textbf{x}_i(t\cdot\Delta t))\right)\right],
\end{equation}
where we used Dirac's delta $\delta$. Similarly, the transition rates are calculated by counting the number of transitions per unit of time. Since the paths in the PE have different weights, the number of transitions is the weighted sum $n_{\mathrm{T}}=\sum_i (h_{\mathrm{AB}}[\textbf{x}_i]+h_{\mathrm{BA}}[\textbf{x}_i])~ w_i$, where the indicator functionals $h_{\mathrm{AB}}$ and $h_{\mathrm{BA}}$ select the $\mathrm{A}$-to-$\mathrm{B}$ and $\mathrm{B}$-to-$\mathrm{A}$ transitions, respectively. Analogously, the total time is $\mathcal{T}=\sum_i (L[\textbf{x}_i] - 1)~w_i~\Delta t$. The rate $k_{\mathrm{AB}}$ is the inverse of the mean first transition time from $\mathrm{A}$, hence we must restrict our sums to in-$\mathrm{A}$ segments and excursions from $\mathrm{A}$ (selected by the functional $h_{\mathrm{A}}$):
\begin{subequations}
\begin{equation}
    k_{\mathrm{AB}} = \frac{\sum_i h_{\mathrm{AB}}[\textbf{x}_i]~ w_i}{\sum_j h_{\mathrm{A}}[\textbf{x}_j]~(L[\textbf{x}_j] - 1)~w_j}~\Delta t^{-1}.
\end{equation}
The rate $k_{\mathrm{BA}}$ counts the in-$\mathrm{B}$ segments and excursions from $\mathrm{B}$ instead:
\begin{equation}
    k_{\mathrm{BA}} = \frac{\sum_i h_{\mathrm{BA}}[\textbf{x}_i]~ w_i}{\sum_j h_{\mathrm{B}}[\textbf{x}_j]~(L[\textbf{x}_j] - 1)~w_j}~\Delta t^{-1}.
\end{equation}
\end{subequations}

\section{\label{sec:methods}Methods}

\begin{figure*}
\centering
\makebox[\textwidth][c]{\includegraphics[clip,width=\textwidth]{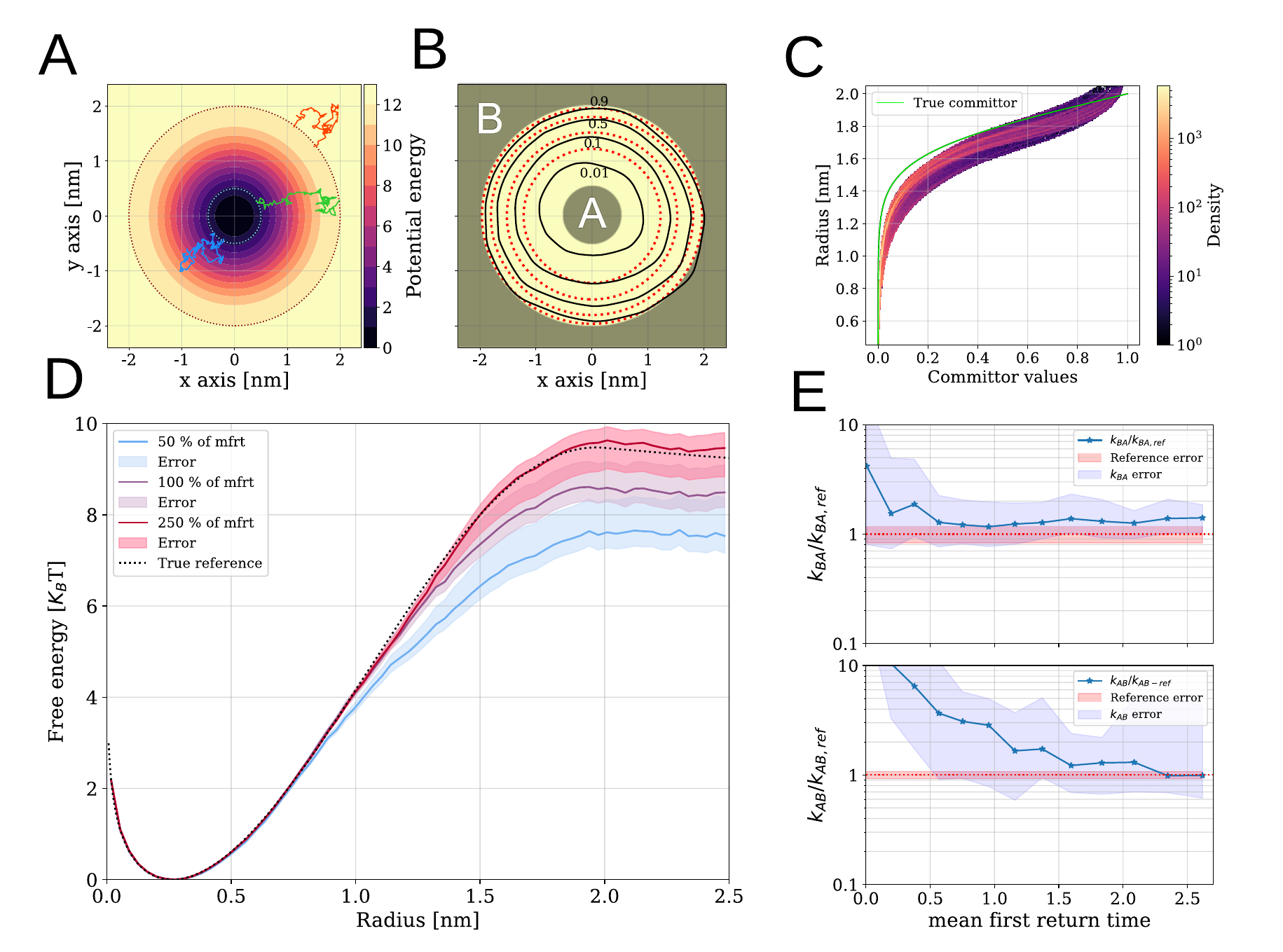}}
\caption{Validation on a 2-dimensional radially symmetric potential. \textbf{A}) $(x,y)$ contour plot of the energy surface. We superimposed representative excursion from state $\mathrm{A}$ (blue), transition (green), and equilibrium segment in $\mathrm{B}$ (red). \textbf{B}) $(x, y)$ contour plot of the latest RC model from AIMMD converted to the expected committor values $p_{\mathrm{B}}(x,y)=\sigma(\lambda(x,y))$ (black lines). For comparison, we also plot the reference committor values $\hat p_{\mathrm{B}}(x,y)$ (red). We highlighted the boundaries of state $\mathrm{A}$ and $\mathrm{B}$. \textbf{C}) The expected committor evaluated on the training set ($y$-axis) versus the reference values $\hat p_{\mathrm{B}}$ ($x$-axis) for all the simulated configurations outside of $\mathrm{A}$ and $\mathrm{B}$. \textbf{D}) Free energy estimates from the reweighted PE projected on the radial coordinate $r$ at levels of cumulative sampling (details in Table 1). Dotted line: analytical free energy. The shaded regions represent the 95\% confidence intervals obtained by bootstrapping on the sampled trajectory segments. \textbf{E}) Evolution of the rates estimates (top: $k_{\mathrm{AB}}$, bottom: $k_{\mathrm{BA}}$) compared to the reference at different stages of AIMMD. The blue-shaded regions are the 95\% confidence intervals obtained by bootstrapping on the sampled trajectory segments. The red-shaded regions are the 95\% confidence interval of the reference estimates.}
\label{fig:2}
\end{figure*}

\begin{figure*}
\centering
\makebox[\textwidth][c]{\includegraphics[clip,width=1.05\textwidth]{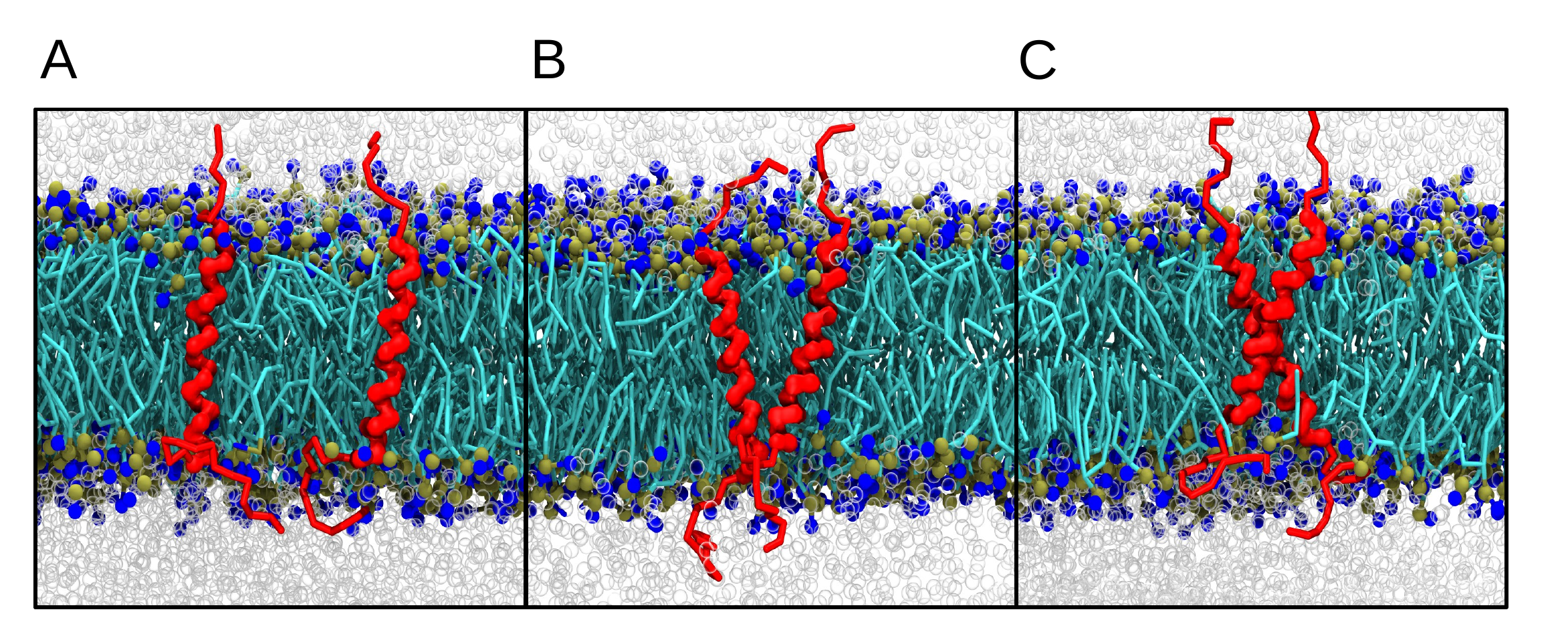}}
\caption{EGFR-ErbB1 transmembrane protein domain dimerization, renders of representative sections of unbound (left), transition state (center), and bound configurations (right) from equilibrium MD simulations. The system was modeled with the Martini3 coase-grained force field. The monomers are in red cartoon representation; the DLPC lipids are shown in Van der Waals representation for the lipid heads beads and licorice representation for the lipid tails, and the solvent is in transparency.}
\label{fig:3}
\end{figure*}

\subsection{\label{sec:toy-system}2-dimensional radially symmetric potential}

We first validated our method on the 2-dimensional radial potential (Fig.~\ref{fig:2}):
\begin{equation}
    V(r) = \begin{cases}
        6~(1-\cos \frac{r}{\pi})~&[k_{\mathrm{B}}T]\quad\text{if }r < 2,\\
        12~&[k_{\mathrm{B}}T]\quad\text{if }r \geq 2,
    \end{cases}
\end{equation}
with $r = \sqrt{x^2+y^2}$. We simulated the dynamics of a particle subject to $V(x,y)$ in a $5\times 5~[\mathrm{L}]^2$ periodic box with an overdamped, isotropic Langevin dynamics~\cite{peters2017reaction}:
\begin{equation}\label{langevin}
\begin{aligned}
    x(t+dt) &= x(t) - \nabla_x V(x,y)~dt + \zeta_x(t)~\sqrt{2Ddt},\\
    y(t+dt) &= y(t) - \nabla_y V(x,y)~dt + \zeta_y(t)~\sqrt{2Ddt},
\end{aligned}
\end{equation}
where $dt$ is the integration step and $D$ the diffusion coefficient ($Ddt = 10^{-5}~[\mathrm{L}]$), and $\zeta_x(t),~\zeta_y(t)$ are sampled from a Normal distribution. We directly implemented Eqs.~\eqref{langevin} in Python. We saved trajectory frames every 100 integration steps during the simulations. We set the state boundaries based on the radius:
\begin{equation}
\begin{aligned}
    \mathrm{A} &= \{(x, y)\mid r \leq 0.5\}\\
    \mathrm{B} &= \{(x, y)\mid r \geq 2.0\},
\end{aligned}
\end{equation}
state $\mathrm{A}$ being around the bottom of the well, and state $\mathrm{B}$ encompassing the interaction less region of the configuration space. We ran an equilibrium simulation for $4\times 10^{10}~[dt]$ steps and detected $338$ transition events, from which we obtained the reference rates $k_{\mathrm{AB}}^{\mathrm{ref}}=(4.3 \pm 0.3)\times 10^{-9}~[dt]^{-1}$ and $k_{\mathrm{BA}}^{\mathrm{ref}}=(2.4 \pm 0.4)\times 10^{-6}~[dt]^{-1}$.

We obtained the reference ``true committor'' by solving numerically the steady state solution of the Fokker-Plank equation for overdamped, isotropic Langevin dynamics~\cite{peters2017reaction}:
\begin{equation}
    (\nabla V(x,y)-D\nabla)\cdot \nabla~p_{\mathrm{B}}(x,y) = 0,
\end{equation}
with boundary conditions $p_{\mathrm{B}}(x,y)\equiv 0,~\forall (x,y)\in\mathrm{A}$ and $p_{\mathrm{B}}(x,y)\equiv 1,~\forall (x,y)\in\mathrm{B}$. Specifically, we applied the relaxation method~\cite{covino_molecular_2019} on a $500\times 500$ grid that converged in less than 30 minutes on a mid-range workstation. We implemented the method in Python in the function ``solve\_committor\_by\_relaxation'', which is available in the ``utils.py'' file of this paper's repository.

We assigned confidence intervals to our estimates by bootstrapping~\cite{diciccio1996bootstrap}. Specifically, we created a new PE instance by selecting random trajectories without replacement from the current PE. These trajectories begin and end in either state $\mathrm{A}$ or $\mathrm{B}$ and have a single weight associated with their shooting point. We then reweighted the PE with the committor model available at that step number and calculated the free energy from the result. After repeating this procedure a hundred times, we calculated the mean free energy value and the standard deviation for every bin. Finally, we plotted the mean value plus/minus the standard deviation of the bootstrapping for every bin as the error of the free energy profile. For the error of the rates, we performed a bootstrapping procedure similar to that of the free energy profile. Instead of calculating the free energy from the new and reweighted PEs, we calculated the dimerization and dissociation rates. For each of the resulting rate coefficients, we took the 2.5 and 97.5 per cent quantiles in the logit space as the lower and upper values of the error.

\subsection{\label{sec:martini-helix}Dimerization of EGFR-ErbB1 TMD in Martini coarse-grained simulations}

We took the crystal structure of the transmembrane domain (TMD) of ErbB1 (PDB id: 2M0B, comprising 13 N-terminal and 9 C-terminal tail residues) and modelled it with the Martini 3 coarse-grained force field~\cite{souza2021martini} (Fig.~\ref{fig:3}). We used Charmm-GUI to convert the atomistic coordinate to Martini beads~\cite{jo2008charmm, qi2015charmm, hsu2017charmm}. Replicating the benchmark of Souza et al.~\cite{souza2021martini}, we put two homodimers in a 100\% DLPC lipid bilayer and solvated the system with a $0.5$~M sodium chloride concentration in a $11.5\times 11.5\times 10.8~\text{nm}^3$ box with periodic boundary conditions. Both the membrane and the solvent were built with the insane Python software~\cite{wassenaar2015computational}. The resulting system consisted of 11796 beads and 408 lipids. We ran all simulations on Gromacs 2023~\cite{abraham2015gromacs} in NPT settings with a 310~K v-rescale thermostat~\cite{bussi2007canonical}, and a 1~atm Parrinello-Rahman semi-isotropic barostat ($\tau_{\mathrm{p}}=12~\text{ns}$)~\cite{quigley2004langevin}. We set an integration step $dt=20~\text{fs}$ and saved trajectory frames every $150~\text{ps}$.

To characterize the system, we chose the distance RMSD (dRMSD) as CV:
\begin{equation}\label{eq:drmsd}
    \mathrm{dRMSD}(x) = \sqrt{\dfrac 1 N~\sum_{(i,~j)\in S}~(d_{ij}(x)-d^0_{ij})^2},
\end{equation}
where $S$ is the set of all the 1763 $(i,~j)$ inter-bead combinations of the transmembrane domains, $d_{ij}(x)$ is the Euclidean distance between those beads for configuration $x$, and $d^0_{ij}$ is the (small) native distance of a reference dimerized structure. The outcome is a CV that is lowest when the monomers are together and grows when the monomers undergo dissociation; in the limit of very far monomers, it converges to the distance between the centers of mass of the individual monomer helices (defined as the interhelical distance). As such, it is a natural choice for setting the boundaries of the metastable states:
\begin{equation}
\begin{aligned}
    \mathrm{A} &= \{x\mid \mathrm{dRMSD}(x) \leq 0.5~\mathrm{nm}\}\\
    \mathrm{B} &= \{x\mid \mathrm{dRMSD}(x) \geq 1.5~\mathrm{nm}\}.
    \end{aligned}
\end{equation}

We ran ten 100~µs-long equilibrium simulations from which we extracted the reference free energy profiles for both the complete path ensemble (equivalent to the Boltzmann distribution), a sample of the transition paths ensemble, and the reference transition rates $k_{\mathrm{AB}}^{\mathrm{ref}}=0.10 \pm 0.01~\text{µs}^{-1}$ and $k_{\mathrm{BA}}^{\mathrm{ref}}=2.6 \pm 0.4~\text{µs}^{-1}$ (Fig.~S1). The association rate depends on the monomers' concentration on the membrane, $C_M$~\cite{sigmundsson2002determination}. If $C_M$ is sufficiently small, the dependence is approximately linear:
\begin{equation}
    k_{\mathrm{BA}} = k_{\mathrm{ass}}~C_M,
\end{equation}
and $k_{\mathrm{ass}}$ is in units of frequency divided by concentration. We can derive $k_{\mathrm{ass}}$ from our system by estimating $C_M$ from the simulation box size. In particular, $C_M \approx 1/(11.5\times 11.5)~\mathrm{nm}^{-2}$, as each monomer would find another every $11.5\times 11.5~\mathrm{nm}^{2}$.

\subsection{\label{sec:aimmd-sim}AIMMD simulations}

We conducted an AIMMD run for each of the systems, using the data production and analysis algorithm available at the repository \url{http://10.5281/zenodo.13145057}. We dedicated two computing nodes to two-way shooting simulations, three nodes to equilibrium simulations around $\mathrm{A}$, and two nodes to equilibrium simulations around $\mathrm{B}$. Every time an equilibrium simulation underwent a transition, we reinitialized it at a random configuration in its target state. Another computing node acts as a ``manager'' by controlling the two-way shooting simulations, collecting the generated trajectories in the PE, and training the ML model. The code requires an initial transition from which to initialize the early simulations, which does not have to be a physical trajectory. For the radial potential, we took a straight line connecting $\mathrm{A}$ and $\mathrm{B}$; for EGFR-ErbB1, we used a $\mathrm{B}$-to-$\mathrm{A}$ transition event observed in a short unbiased run from the dissociated state.

The RC model $\lambda(x)$ is obtained through simple feed-forward NNs implemented in pytorch~\cite{paszke2019pytorch}. The network of the radial potential receives the $(x, y)$ positions as input features and has a $(2,~512,~512,~512,~1)$ architecture with PReLU activation functions except for the linear output layer. The network of ErbB1 receives all the 1763 inter-bead distances in the dRMSD computation of Eq.~\ref{eq:drmsd} as input features. It has a $(1763,~1000,~1000,~1000,~1)$ architecture with ELU activation functions except for the linear output layer (Fig.~S3). At regular AIMMD steps, corresponding to the completion of a two-way shooting simulation, we re-trained the networks from scratch with a loss function based on the outcomes of the simulations. The training set has as many elements as the simulated excursions and transitions. It is composed by the $(x_{\mathrm{sp},i}, r_{\mathrm{A},i}, r_{\mathrm{B},i})$ triplets, where $x_{\mathrm{sp},i}$ is the trajectory's shooting point if it comes from a two-way shooting simulation or a random frame of the trajectory if it comes from an equilibrium simulation, $r_{\mathrm{A},i}$ is the number of times the trajectory crosses the boundary of $\mathrm{A}$, and $r_{\mathrm{A},i}$ is the number of times the trajectory crosses the boundary of $\mathrm{B}$. Before training, we weighted the training set elements such that we had a uniform shooting point distribution in the $\lambda \approx \sigma^{-1}(p_{\mathrm{B}})$ space. Then, we minimized the loss~\cite{peters2010recent,jung2023machine}
\begin{equation}
    \mathcal L = \sum_{i \in \text{batch}} \left[r_{\mathrm{A},i}~\log \sigma(-\lambda(x_{\mathrm{sp}, i}))+r_{\mathrm{B},i}~\log \sigma(\lambda(x_{\mathrm{sp}, i}))\right]
\end{equation}
over 500 training epochs, where we resampled the training set with repetition in batches of 4096 elements every time. The learning rate was $l_r=10^{-4}$ with the Adam optimizer (radial potential) or $l_r=3\times 10^{-4}$ with the AdamW optimizer~\cite{loshchilov2017decoupled} (weight decay $w_d=0.06$, EGFR-ErbB1). We optimized the EGFR-ErbB1 hyperparameters \textit{ex-post} on preliminary AIMMD data (Fig.~S2). In general, it is always possible to improve the model after the sampling campaign has ended, leading to improved free energies and rate estimates. The ML part takes only a small fraction of the time required by the simulations.

To estimate the crossing probabilities of $P_{\mathrm{A}}(\lambda),~P_{\mathrm{B}}(\lambda)$, we need to consider both equilibrium and two-way shooting excursions. However, suboptimal RC models can negatively affect these estimates, leading to systematic errors that inflate the transition rate estimates. These errors are largest near the metastable states. To minimize these errors, we set a threshold $\lambda_{\mathrm{A}}$, which is the farthest RC interface from $\mathrm{A}$ such that at least $n_{\mathrm{eq}}=6$ equilibrium excursions reach $\lambda_{\mathrm{A}}$. Only the equilibrium excursions from $\mathrm{A}$ contribute to the estimate of $P_{\mathrm{A}}(\lambda)$ up to $\lambda_{\mathrm{A}}$. Similarly, only the equilibrium excursions from $\mathrm{B}$ contribute to the estimate of $P_{\mathrm{B}}(\lambda)$ up to $\lambda_{\mathrm{B}}$. As the AIMMD run continues and more equilibrium excursions are simulated, $\lambda_{\mathrm{A}}$ is pushed closer to $\mathrm{B}$ and $\lambda_{\mathrm{B}}$ is pushed closer to $\mathrm{A}$. With infinite sampling, we can recover the crossing probability and rate estimates that we would observe in a long unbiased MD simulation since only the equilibrium simulations would contribute to those estimates. The two-way shooting simulations would still add to the free energy estimate in the transition region. AIMMD speeds up convergence in the data-scarce regime, where we do not have enough data from equilibrium simulations alone to provide meaningful estimates.

\begin{table}[h]
\caption{\label{tab:table1}Summary of all simulations.
}
\begin{ruledtabular}
\begin{tabular}{lccccccc}
System&
\begin{tabular}{c} 2-way\\shots\end{tabular} &
\begin{tabular}{c} Transition\\paths\end{tabular} &
\begin{tabular}{c} Equilibrium~A\\excursions\end{tabular} &
\begin{tabular}{c} Equilibrium~B\\excursions\end{tabular} &
\begin{tabular}{c} Simulation\\time $[\mathrm{mfrt}]$\end{tabular} &
$\dfrac{k_{\mathrm{A}\mathrm{B}}}{k_{\mathrm{A}\mathrm{B}}^{\mathrm{ref}}}$&
$\dfrac{k_{\mathrm{B}\mathrm{A}}}{k_{\mathrm{B}\mathrm{A}}^{\mathrm{ref}}}$
\vspace{2pt}
\\
\colrule
2D radial & 1348 & 385 & 11448 & 461 & 0.51 & $5.7 \pm 1.6$ & $1.3 \pm 0.7$ \\
          & 2650 & 754 & 22561 & 820 & 1.02 & $2.1 \pm 0.8$ & $1.2 \pm 0.8$ \\
          & \textbf{5806} & \textbf{1642} & \textbf{56926} & \textbf{1682} & \textbf{2.52} & $\mathbf{0.9} \pm \mathbf{0.6}$ & $\mathbf{1.4} \pm \mathbf{0.5}$ \\ \colrule
EGFR-ErbB1 & 212 & 87 & 1329 & 214 & 1.51 & $2.0 \pm 1.2$ & $1.5 \pm 0.7$ \\
          & \textbf{546} & \textbf{214} & \textbf{3676} & \textbf{347} & \textbf{3.55} & $\mathbf{1.2} \pm \mathbf{0.8}$ & $\mathbf{1.3} \pm \mathbf{0.6}$
\end{tabular}
\end{ruledtabular}
\end{table}

\section{\label{sec:results}Results}

\subsection{Analytical benchmark system}

We validated our algorithm using a particle diffusing in a 2-dimensional analytical potential that recapitulates the features of transmembrane dimerization (Fig.~\ref{fig:2}). The radially symmetric potential has a single central deep well in a flat surface. It models dimerization seen from one of the helices placed at the center of the potential. The  12 $k_{\mathrm{B}}T$ deep well represents the bound state, whereas the flat surface around it represents the configuration space describing the unbound dimer. The particle does not experience any barrier before falling into the well. The simplicity of this potential allows for free energy and rate estimates through equilibrium simulations. Also, we could numerically solve the Fokker-Plank equation and obtain a reference value of the committor (see Section~\ref{sec:toy-system}).

AIMMD samples short trajectory segments on the potential and learns the committor associated to the process on the fly. Trajectory segments are obtained either by simulating short equilibrium trajectories in states $\mathrm{A}$ and $\mathrm{B}$ or by path sampling in the transition region (Fig.~\ref{fig:2}A). Path sampling can result in excursions from one of the two states or transitions connecting them. At convergence, AIMMD learns an accurate model of the committor, which, in turn, allows for optimal production of transition paths at around 30\% of the total path sampling simulations (Table~\ref{tab:table1}). Importantly, all simulated segments are very short, with the exception of the in-$\mathrm{B}$ segments that are the outcome of easy-to-model Brownian diffusion in a box. We quantified the total computational cost by reporting the cumulative simulated time in units of mean first return time (mfrt), i.e., the average time it takes for the system to go from A to B or vice versa:
\begin{equation}
    \mathrm{mfrt} = \tau_{\mathrm{AB}}+\tau_{\mathrm{BA}} = \frac{1}{k_{\mathrm{AB}}^{\mathrm{ref}}}+\frac{1}{k_{\mathrm{BA}}^{\mathrm{ref}}},
\end{equation}
where $\tau_{\mathrm{AB}}$ and $\tau_{\mathrm{BA}}$ are the mean first passage time from $\mathrm{A}$ to $\mathrm{B}$ and vice-versa, respectively~\cite{peters2017reaction}. We calculated free energy and rates as a function of increasing computational budget, from approx. 0.5 to 2.5 mfrt (Table~\ref{tab:table1}).    

Despite the challenge of learning the committor from actual simulations, we obtained a sufficiently accurate model. Figure~\ref{fig:2}B compares the isolines of the committor $p_{\mathrm{B}}=\sigma(\lambda)$ learned by AIMMD from path sampling simulations with a direct solution of the Fokker-Planck equation. The committor is most accurate around the TSE, where $p_{\mathrm{B}}\sim 0.5$, whereas it is the least accurate close to the state boundaries. This is expected from shooting points-based training: fine-tuning the committor at different isolines requires both $n_{r_{\mathrm{A}}} > 1$ and $n_{r_{\mathrm{B}}} > 1$, which is exponentially more difficult approaching the metastable states. For very large energy barriers, the committor between two states is approximately a step function, almost 0 or 1 close to the states, and with a steep gradient only in a small region of the reactive space. Learning the committor accurately becomes exponentially difficult. Notably, even though the learned committor isolines are only approximately accurate, they never cross the actual ones. This means they correctly sort different points along the progression of the $\mathrm{A}$ to $\mathrm{B}$ transition, allowing for accurate reweighting.

Our benchmark potential also allows us to compare the radial distance $r$---the obvious RC choice for this system---with the actual committor. A two-dimensional histogram of all sampled configurations as a function of their committor (as predicted by the last NN model) and the value of radial coordinates show the functional mapping between the two coordinates. The plot also shows the functional relationship between the radial coordinate and the ground truth committor.
The radial coordinate is a linear function of the committor almost everywhere. However, $r$ is a poor parametrization of the committor close to state $\mathrm{A}$. In summary, this analysis suggests that the radial coordinate is a good proxy of the committor but fails to resolve the mechanism very close to the bound state. 

After restoring the equilibrium weights for all trajectory segments, we can project the weight of all sampled configurations along an arbitrary variable to obtain a free energy profile. In this case, the obvious choice was the radial coordinate $r$. While a cumulative sampling corresponding to 50\% of the mfrt is sufficient for a rough estimate of the free energy profile, using the equivalent of 205\% of the mfrt provides a very accurate estimate.

We also obtain very good estimates of the transition rates with a limited amount of sampling. The rate of the fast process in the system, corresponding to the particle falling in the well, i.e., the dimer assembling, converges already after only 0.5 mfrt of sampling. The slower rate, corresponding to the particle leaving the well, requires instead approximately 1.5 mfrt to converge. However, very little sampling is enough to provide a within-an-order-of-magnitude estimate of both rates. For comparison, the reference transition rates (of around 80\% accuracy) were obtained by using 170 mfrt of equilibrium sampling (totaling 340 transitions). 

\begin{figure*}
\centering
\makebox[.\textwidth][c]{\includegraphics[width=\textwidth]{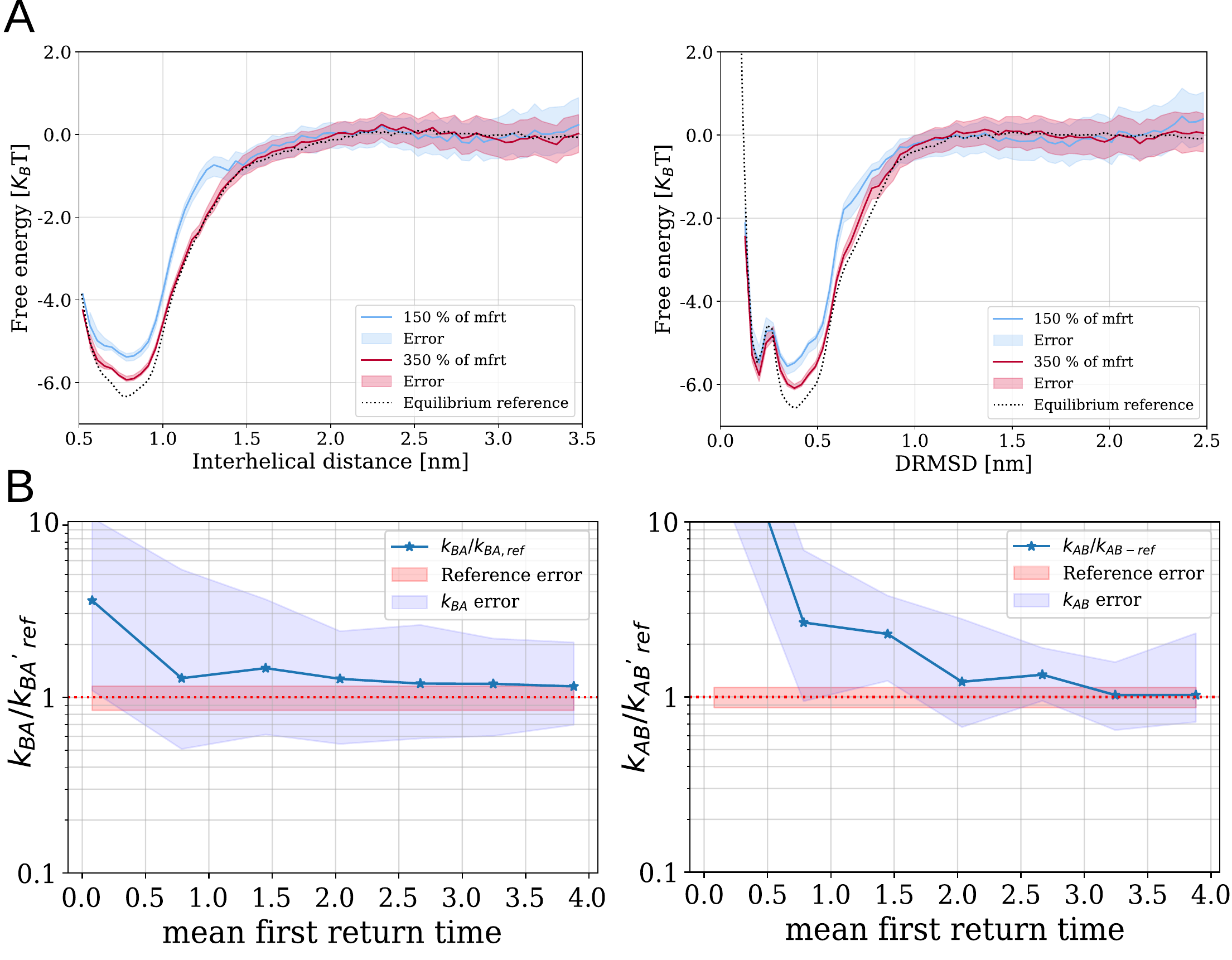}}
\caption{EGFR-ErbB1 transmembrane protein domain dimerization, results of the AIMMD run. \textbf{A}) Free energy estimates from the reweighted PE projected on the interhelical distance (left) and on the dRMSD CVs (right) at different stages of AIMMD. Dotted line: free energy from 1~ms long reference MD simulations. The shaded regions represent the 95\% confidence intervals obtained by bootstrapping on the sampled trajectory segments. 
Evolution of the rates estimates (left: $k_{\mathrm{BA}}$, right: $k_{\mathrm{AB}}$) compared to the reference at different stages of AIMMD (details in Table 1). The blue shaded regions are the 95\% confidence intervals obtained by bootstrapping on the sampled trajectory segments. The red-shaded regions are the 95\% confidence interval of the reference estimates.}
\label{fig:4}
\end{figure*}

\begin{figure*}
\centering
\makebox[\textwidth][c]{\includegraphics[width=.48\textwidth]{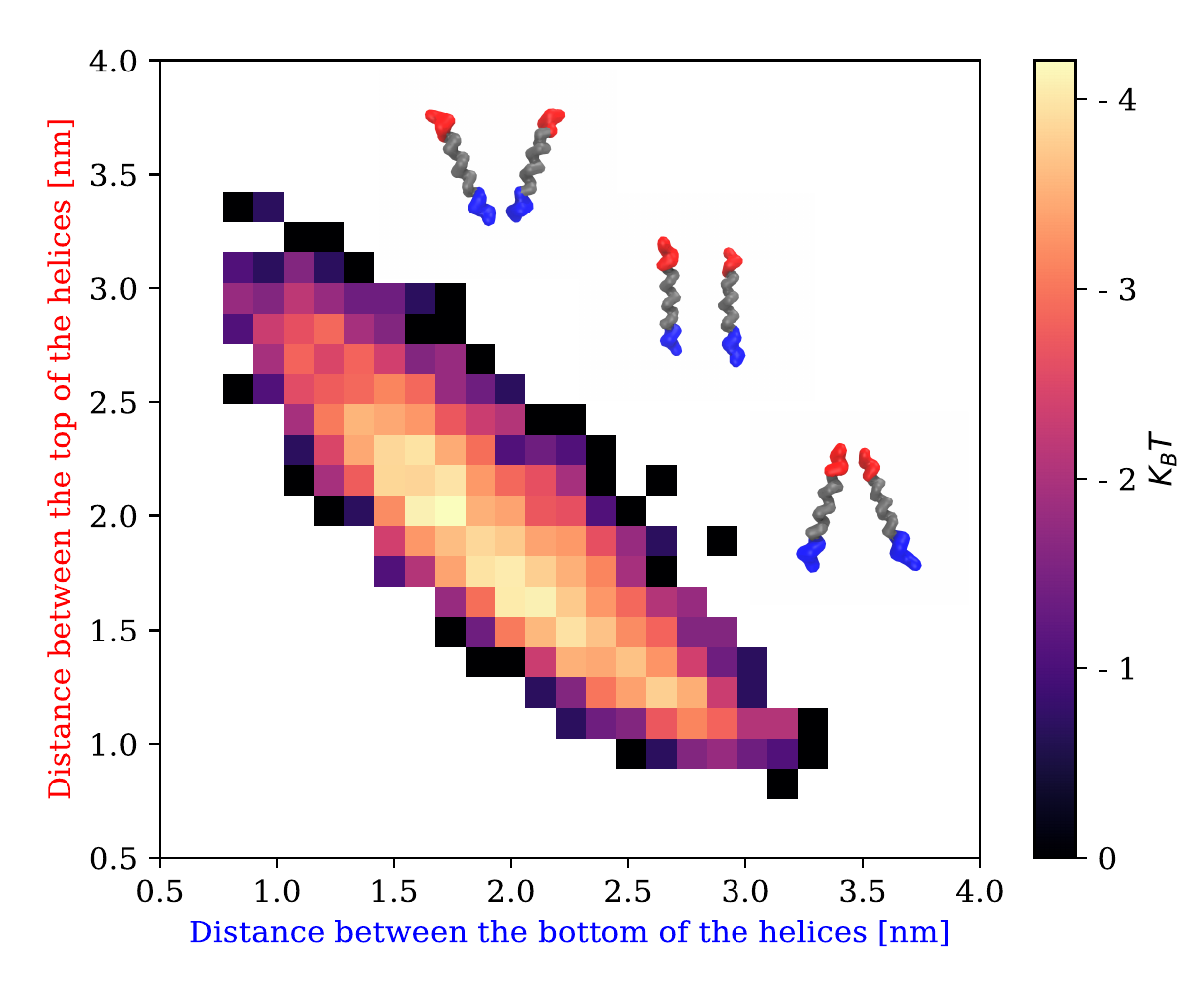}}
\caption{EGFR-ErbB1 transmembrane protein domain dimerization, free energy of the transition state projection on the $(d_1,~d_2)$ space, where $d_1$ is the distance between the bottoms of the transmembrane helices, and $d_2$ is the distance between the tops. Superimposed renders of representative dimers configurations. Only the transitions paths configurations with committor model $p_{\mathrm{B}}\in [0.4,0.5]$ where plotted. The top-left (approaching from the top), center (approaching in a parallel way), and bottom-right (approaching from the bottom) show the three possible dimerization pathways, all explored by the AIMMD sampling.}
\label{fig:5}
\end{figure*}

\subsection{Transmembrane dimerization of EGFR-ErbB1 homodimers in a lipid bilayer}

We applied our algorithm to study the dimerization of the EGFR-ErbB1 transmembrane domain. The ErbB family contains receptor tyrosine kinases that have a crucial role in controlling the growth and survival of cells. Aberrant activation of these receptors can lead to a constitutively active or even hyperactive kinase and favor tumor progression~\cite{trusolino2002scatter, yarden2012erbb, hendler1984human, wheeler2015receptor}. Because of the high expression levels and central role in tumors, these receptors have been an attractive target for cancer therapy. The development of these methods would benefit from a mechanistic understanding of the activation process, which relies on transmembrane dimerization.

Given its importance as a drug target, ErbB1 has been widely characterized, making it an excellent benchmark for studying transmembrane dimerization. The structure of its native dimer and the free energy of dimerization are experimentally known. Additionally, the TMD homodimer of ErbB1 spontaneously dissociates in long Martini 3 equilibrium simulations (Fig.~S1). Thus, we could estimate ground truth reference values for the free energy and rates from 1~ms of total equilibrium simulations (approx. 95 mfrt), making it an ideal validation system for our algorithm.

We efficiently estimated the PE characterizing ErbB1's dimerization with AIMMD by sampling many short trajectory segments. We modeled a system containing two transmembrane helices in a solvated bilayer of DLPC lipids in Martini 3. We estimated free energy and rates for increasing values of cumulative sampled time, approx. 1.5 and 3.5 mfrt, where 1 mfrt $\approx 10.5~$µs (Fig.~\ref{fig:3} and Table~\ref{tab:table1}). In both cases, we obtained a production rate of transitions around 40\%, consistent with an accurate committor model learned by the NN. The dRMSD time series of a representative transition path with snapshots of the dimer is shown in Fig.~S4.

A limited amount of cumulative sampling is sufficient to estimate the free energy of dimerization accurately. After reweighting all the trajectory segments, we could estimate a free energy profile along an arbitrary order parameter or collective variable by projecting and calculating a histogram. Figure~\ref{fig:4}A shows the profiles as a function of the interhelical distances and DRMSD calculated at two increasing levels of cumulative sampling. A comparison with the free energy profile extracted from long equilibrium runs shows that our estimates are accurate and can also resolve fine features of the profiles. In particular, the free energy barrier height is within 0.5 $k_{\mathrm{B}}T$ of the reference. It is important to stress that all trajectory segments were sampled without using a guiding force defined along collective variables.

Without any additional calculations, we also obtained accurate transition rates estimates within a factor two of the reference and characterized the TSE. Like in the benchmark system, estimating the association rate required less sampling than the dissociation rate, which is the slow process in this system. The trained NN approximating the committor is a function of the system's configuration that is quick to evaluate, enabling us to identify the TSE as defined by the set of configurations with a committor value between 0.4 and 0.6. Figure~\ref{fig:5} shows a two-dimensional free energy of the TSE as a function of the distance between the bottom of the TMHs and the top, respectively. The TSE shows that our algorithm could sample a heterogeneous dimerization mechanism. In most cases, the TMHs associate simultaneously, whereas top-to-bottom and bottom-to-top zipping associations are also possible. 

\section{\label{sec:discc}Discussion}

Characterizing the assembly of membrane proteins in lipid bilayers is of outstanding biological interest. In this paper, we propose a novel algorithm to sample the assembly of TMHs and obtain accurate estimates of free energies, rates, and the dimerization mechanism. Fully resolving the assembly of the EGFR-ErbB1 homo-dimer in Martini 3 simulations required less than 40 µs, a very modest computational cost.

Most approaches that estimate the energetics of transmembrane protein association use CV-based biased simulations. While these have provided fundamental contributions, obtaining well-converged and accurate free energy profiles for processes in a condensed phase is non-trivial. The reason is that these methods require formulating a CV---low-dimensional function of the system's configuration space---that captures all the slow degrees of freedom. Missing those degrees of freedom will result in an inaccurate (but still precise!) free energy profile. The CV should not only accurately describe the dynamics along the transition mechanism but also, at the same time, resolve the metastable state. Given that transmembrane dimers sample alternative configurations in their bound state, this is an extra challenge that compounds on an already highly non-trivial problem. Current strategies to overcome this challenge focus on identifying and adding these slow degrees of freedom to the CV or using ML to learn from data complex many-body terms. 

The main advantage of our approach is that it relies only on short, dynamically unbiased trajectories. We do not use biasing forces or need to introduce a CV to guide our simulations. AIMMD relies on judiciously initialized short trajectories that maximize the probability of observing spontaneous dimerization events. The reweighting scheme is only necessary to restore these trajectories' correct frequency without reweighing their dynamics. In this way, we can significantly speed up the sampling of the rare dissociation events. Since we do not guide along a prescribed CV and the system evolves according to its unbiased dynamics, our algorithm samples alternative dimerization mechanisms (Fig.~\ref{fig:5}). Also, we decouple sampling in the states from sampling in the transition region, which bears two clear advantages. First, we only need to define the boundary of the states by using any suitable CV or order parameter that helps us to quantify what we mean by ``bound'' and ``unbound.'' Crucially, these coordinates do not have to resolve the dynamics within the state, which we sample with unbiased simulations. Second, we can parallelize the sampling by dedicating different compute nodes to simulations within the states and path-sampling simulations in the transition region.   

Equilibrium sampling internal and close to the metastable states is crucial for accurate rate estimates, as it provides fluxes through the interfaces between the states and the transition region necessary to merge all short trajectories in a single free energy profile and to estimate the rates. The sampling in the states equilibrates independently from the sampling in the transition state. Thus, free energy profiles in the states could be accurate while poorly converging in the transition region and vice-versa. If there are clear substates in the states (like in the ErbB1 bound state around $\mathrm{dRMSD} = 0.2$), then equilibration can be too slow, and a static external bias might be added.  
Sampling the large entropic unbound state can be challenging. Modeling the diffusion in the unbound state as a Brownian process and using the predicted mean first passage time and probability of entering the transition region could replace explicit sampling.

Future work will focus on extending our algorithm to simulations of transmembrane dimerization in membranes with complex lipid compositions and at atomistic resolution. Biological membranes are composed of different lipids that laterally organize in short-lived nanodomains, which profoundly impact the energetics of membrane proteins. The challenge will be to sample efficiently the coupled mixing dynamics of the bilayer and the interactions between the different lipid species and the proteins. Solving this problem will benefit from including the lipids in the input of the NN that approximates the committor function in AIMMD. Additionally, one could use symbolic regression to extract an explicit closed form approximating the committor from the trained NN~\cite{jung_machine-guided_2023}, which could inform more accurate data-driven CVs and provide important mechanistic understanding. Path simulations of transmembrane dimerization at atomistic resolution are feasible, but the very long timescales will make convergence difficult.  

In conclusion, our algorithm offers a promising strategy to study the assembly of membrane proteins in lipid bilayers. By overcoming some of the outstanding challenges of established CV-based sampling strategies, it opens up new avenues for estimating free energy and rates of this key biological process.

\begin{acknowledgments}

E.J., G.L., and R.C. acknowledge Elena Spinetti for fruitful discussions and technical assistance,
the support of Goethe University Frankfurt, the Frankfurt Institute of Advanced Studies, the LOEWE Center for Multiscale Modelling in Life Sciences of the state of Hesse, the CRC 1507: Membrane-associated Protein Assemblies, Machineries, and Supercomplexes (P09), and the International Max Planck Research School on Cellular Biophysics, as well as computational resources and support from the Center for Scientific Computing of the Goethe University and the Jülich Supercomputing Centre.

\end{acknowledgments}

\section*{Data and Code Availability Statement}

All code, data, and analysis scripts needed to reproduce the results in this manuscript are freely and openly available in the Zenodo repository at the web address \url{http://10.5281/zenodo.13145057}. We used custom Python code based on the pytorch\cite{paszke2019pytorch,imambi2021pytorch} (for machine learning) and mdtraj\cite{mcgibbon2015mdtraj} packages (for molecular dynamics trajectories analysis and manipulation). The analysis is streamlined in two JupyterLab notebooks (one for each system).

\bibliography{main}

\end{document}


\title[Supplementray Information]{Supplementary Information for \textit{Free energy, rates, and mechanism of transmembrane dimerization in lipid bilayers from dynamically unbiased molecular dynamics simulations}}
\author{Emil Jackel}
\thanks{Equal contributions}
\affiliation{Institute of Biophysics, Goethe\ University Frankfurt, Frankfurt am Main, Germany.}
\affiliation{Frankfurt Institute for Advanced Studies, Frankfurt am Main, Germany.}
\author{Gianmarco Lazzeri}
\thanks{Equal contributions}
\affiliation{Institute of Biochemistry, Goethe University Frankfurt, Frankfurt am Main, Germany.}
\affiliation{Frankfurt Institute for Advanced Studies, Frankfurt am Main, Germany.}
\author{Roberto Covino*}
\affiliation{Institute of Computer Science, Goethe University Frankfurt, Frankfurt am Main, Germany.}
\affiliation{Frankfurt Institute for Advanced Studies, Frankfurt am Main, Germany.}
\email[Author to whom any correspondence should be addressed: ]{covino@fias.uni-frankfurt.de}
\date{\today}

\maketitle

\section*{Supplementary figures}

\begin{figure*}[h]
\renewcommand{\thefigure}{S1}
    \centering
    \makebox[\textwidth][c]{\includegraphics[width=\textwidth]{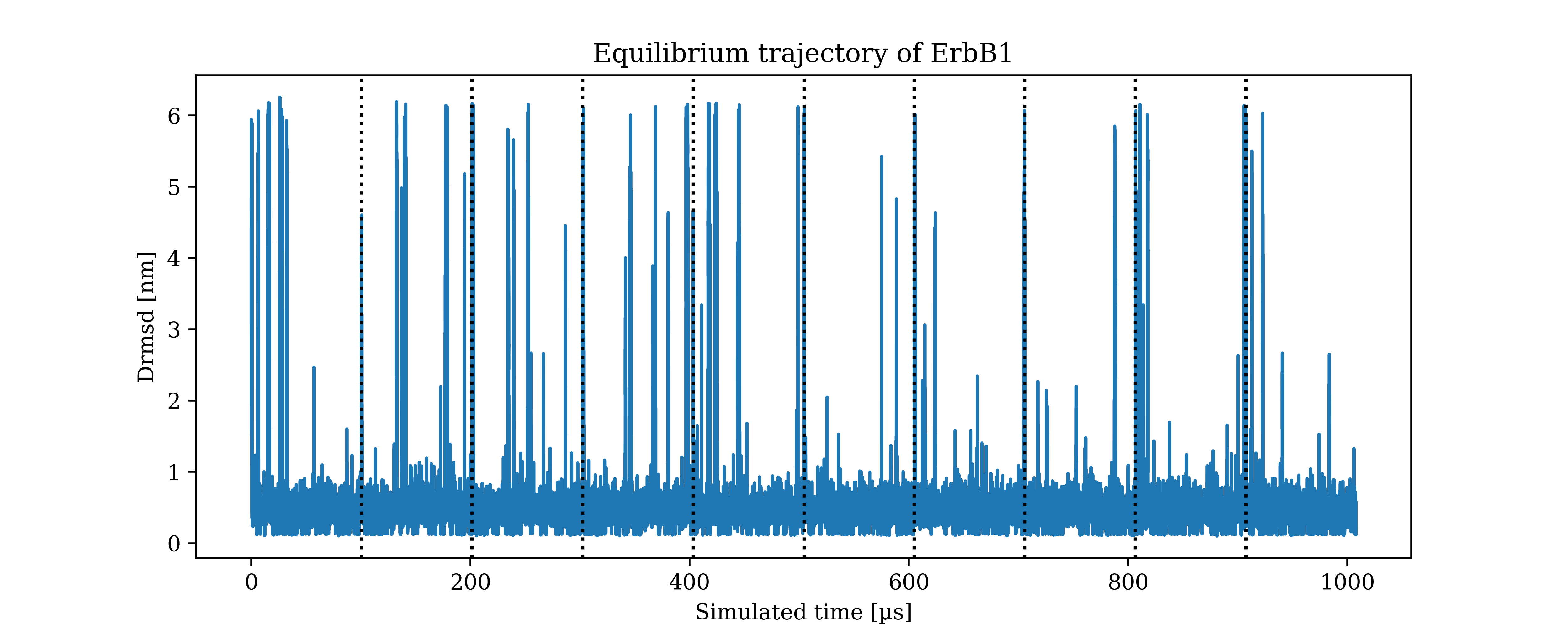}}
    \caption{EGFR system, dRMSD time series of ten 100 µs-long equilibrium simulations replicas for the computation of the reference free energy profiles and rates. Each run is separated from the others by the vertical dotted lines.}
    \label{fig:S1}
\end{figure*}

\begin{figure*}[h]
\renewcommand{\thefigure}{S2}
    \centering
    \makebox[\textwidth][c]{\includegraphics[width=.67\textwidth]{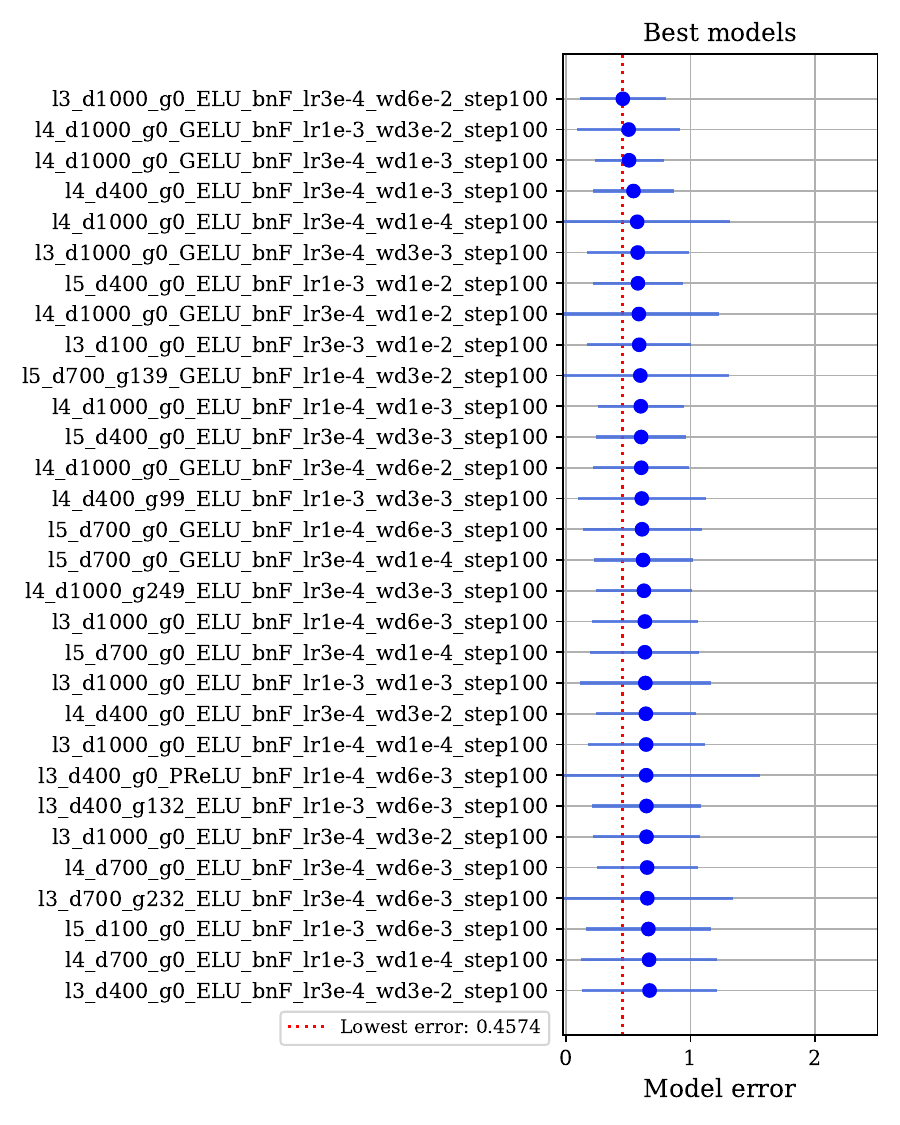}}
    \caption{Machine learning hyperparameters optimization for the RC model for the EGFR system, top 30 results ranked by rescaled RMSE loss on an 87-point test set. We trained 3000 of feed-forward neural network architectures with the AdamW optimizer; the training set came from a 10 µs-long preliminary AIMMD run (100 2-way-shooting simulations). We performed 30 shooting simulations for each point $x_{i,~sp}$ in the test set and estimated its committor value $\hat p_i$. We compared the model's $p_i\equiv \sigma(\lambda(x_{i,~sp}))$ and reference committor estimates and computed the rescaled residuals $e_i = (p_i - \hat p_i) / \min(\hat p_i, 1- \hat p_i)$. The rescaling increases the importance of being accurate close to the metastable states, which in turn improves the performance of the trajectories reweighting algorithm. The best model has the lowest $\mathrm{RMSE}=\sqrt{(\sum_{i=1}^30 e_i^2)/30}$; it has 3 hidden layers of size 1000, ELU activation function, and is trained for 500 epochs with $l_r=3\times 10^{-4}$ and a weight decay of $6\times 10^{-2}$.}
    \label{fig:S1}
\end{figure*}

\begin{figure*}[h]
\renewcommand{\thefigure}{S3}
    \centering
    \makebox[\textwidth][c]{\includegraphics[width=.8\textwidth]{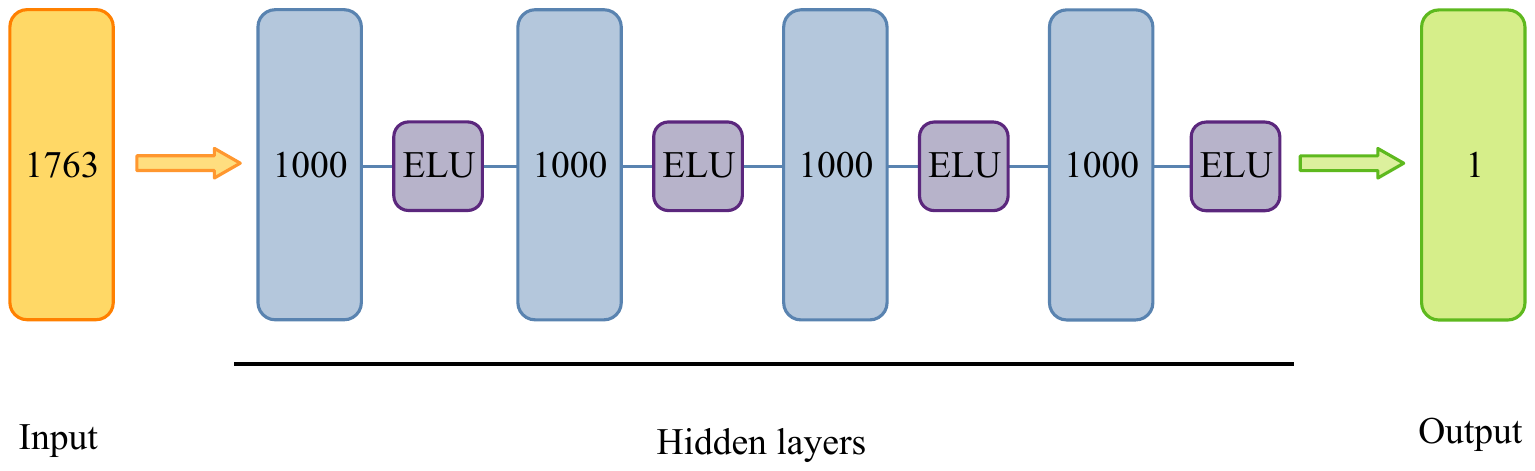}}
    \caption{Neural network architecture of RC model of the EGFR system.}
    \label{fig:S3}
\end{figure*}

\begin{figure*}[h]
\renewcommand{\thefigure}{S4}
    \centering
    \makebox[\textwidth][c]{\includegraphics[width=.67\textwidth]{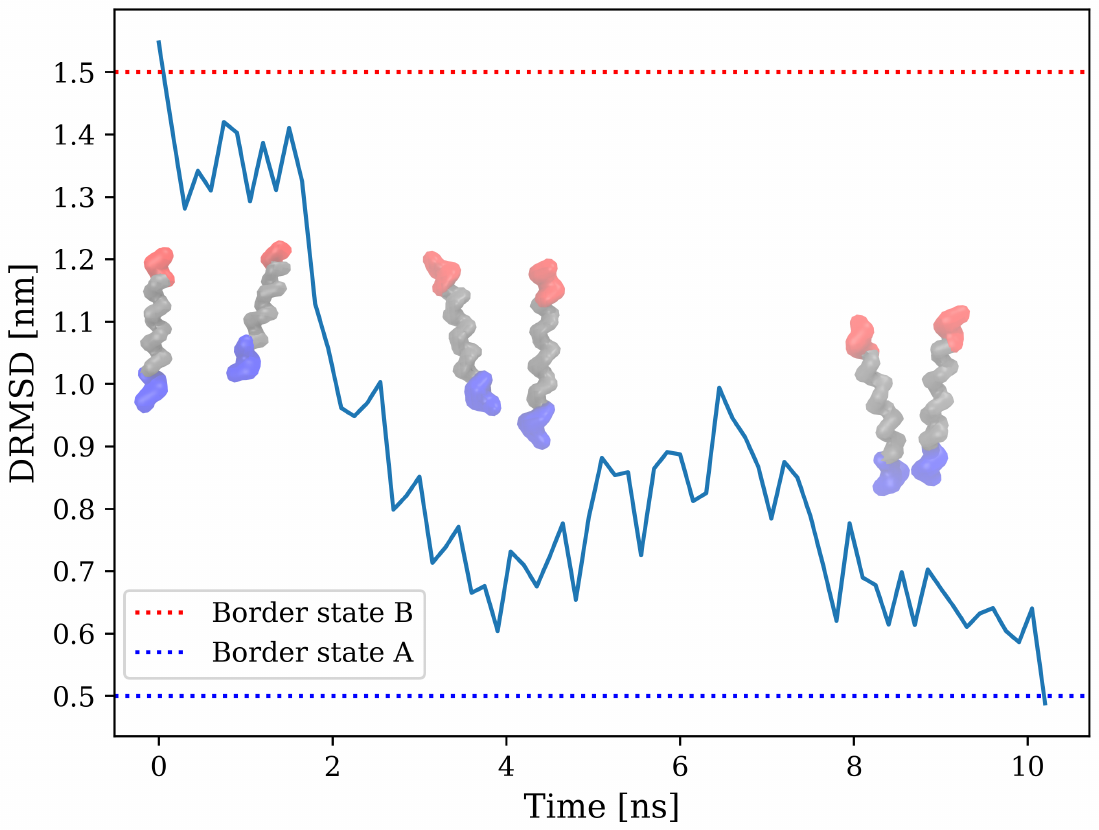}}
    \caption{EGFR, representative transition path from AIMMD, dRMSD time series, with superimposed renders of representative dimer configurations.}
    \label{fig:S5}
\end{figure*}
